\documentclass[preprint, number]{elsarticle}

\usepackage{algpseudocode}
  \usepackage{algorithm}  
  \usepackage{algpseudocode}  
  \usepackage{amsfonts}
  \usepackage{appendix}
  \usepackage{amsmath}
  \usepackage{amssymb}
  \usepackage{amsthm}
  \usepackage{array}
  \usepackage{bm}
  \usepackage{booktabs}
  \usepackage{caption}
  \usepackage{color}
  \usepackage{enumerate}
  \usepackage{float}
  \usepackage{hyperref}
  \theoremstyle{definition}
\newtheorem{definition}{Definition}[section]
\newtheorem{proposition}{Proposition}[section]
\newtheorem{theorem}{Theorem}[section]

\newtheorem{lemma}[theorem]{Lemma}

\begin{document}
\begin{frontmatter}

\title{Enhancing the SVD Compression}

\author[a,b]{Huiwen Wang}
\author[a,c,d]{Yanwen Zhang} 
\author[a]{Jichang Zhao\corref{mycorrespondingauthor}}
\cortext[mycorrespondingauthor]{Corresponding author:}
\ead{jichang@buaa.edu.cn}

\address[a]{School of Economics and Management, Beihang University, Beijing 100191, China}
\address[b]{Key Laboratory of Complex System Analysis, Management and Decision (Beihang University), Ministry of Education}
\address[c]{Beijing Key Laboratory of Emergence Support Simulation Technologies for City Operations, Beijing 100191, China}
\address[d]{Shen Yuan Honors College, Beihang University, Beijing 100191, China}

\begin{abstract}
Orthonormality is the foundation of matrix decomposition. For example, Singular Value Decomposition (SVD) implements the compression by factoring a matrix with orthonormal parts and is pervasively utilized in various fields. Orthonormality, however, inherently includes constraints that would induce redundant degrees of freedom, preventing SVD from deeper compression and even making it frustrated as the data fidelity is strictly required. In this paper, we theoretically prove that these redundancies resulted by orthonormality can be completely eliminated in a lossless manner. An enhanced version of SVD, namely E-SVD, is accordingly established to losslessly and quickly release constraints and recover the orthonormal parts in SVD by avoiding multiple matrix multiplications. According to the theory, advantages of E-SVD over SVD become increasingly evident with the rising requirement of data fidelity. In particular, E-SVD will reduce 25\% storage units as SVD reaches its limitation and fails to compress data. Empirical evidences from typical scenarios of remote sensing and Internet of things further justify our theory and consistently demonstrate the superiority of E-SVD in compression. The presented theory sheds insightful lights on the constraint solution in orthonormal matrices and E-SVD, guaranteed by which will profoundly enhance the SVD-based compression in the context of explosive growth in both data acquisition and fidelity levels.
\end{abstract}

\end{frontmatter}

\section{Introduction}
\label{sec:intro}

In the development history of matrix theory, one of the most fruitful field is matrix decomposition. Of those methodologies, Singular Value Decomposition (SVD), a method factorizing matrix \bm{$X$} into \bm{$U \Sigma V^T$}, in which \bm{$U$} and \bm{$V$} are orthonormal matrices and $\bm{\Sigma} = diag(\sigma_1, \sigma_1,...\sigma_r)$ is a diagonal matrix, has been fully developed both theoretically and practically\cite{stewart1993early}.

As early as $19^{th}$ century, Beltrami~\cite{beltrami1873sulle} firstly derived SVD with bilinear form, Jordan~\cite{jordan1874memoire} completed this method with an elegant derivation, and Sylvester~\cite{sylvester1889reduction} described an iterative algorithm to accelerate the diagnolization procedure. In the $20^{th}$ century, Schimidt~\cite{schmidt1908theorie} introduced the infinite-dimensional analogue of SVD, which extended its application area from linear algebra to integral equations. Schmidt\cite{schmidt1908theorie} further proposed an approximation theorem, proving that SVD can be used to obtain the optimal low-rank approximation of an operator. Beyond these classic work, SVD theories continue to expand in recent decades \cite{Golub1971, de2000multilinear, grasedyck2010hierarchical, zhang2019optimal, kilmer2021tensor}, and theoretical works with application purpose are also actively studied especially in recent years \cite{szalontai2020svd, doi:10.1126/sciadv.aay7170, nadi2021optimal, delannoy2021evaluating}.

These advances in both theory and practice give rise to the really special role of SVD in matrix decomposition and have attracted lots of research interests. In particular, some of its theoretical advantages makes it really powerful for compression \cite{scharf1991svd}: firstly and most importantly, for any matrix \bm{$X$} whose rank is $r$, if we choose a specific positive integer $l<r$, SVD is able to figure out the optimal $l-$rank approximation matrix $\bm{\hat{X}}$, namely minimizes the sum of the squares of the difference between elements of \bm{$X$} and \bm{$\hat{X}$}, i.e, $\sum_{i,j}|x_{ij} - \hat{x}_{ij}|^2$. Moreover, efficient and solid algorithms of computing various types of SVD have been pervasively established, and mature packages are available in almost all programming software. Therefore, SVD as an efficient compression technique has been widely applied in many fields.

The first significant application domain of SVD compression is digital images, which are broadly collected from areas such as biology \cite{fan2021bioorthogonal}, medicine \cite{glatard2012virtual}, agriculture \cite{10.1093/nsr/nww015}, astronomy \cite{janssen2021event} and remote sensing \cite{kuhn2021declining}. With the rapid increase of the number and size of images, compression becomes indispensable and urgent to get less storage space as well as faster transmission speed \cite{xin2021lossless}. SVD is suitable for this task due to that images can be reconstructed by fewer amounts of singular values~\cite{andrews1976singular}. Compression by SVD shows high perceptual quality, leading many researches continue to develop it and combine it with other methods to get better performance \cite{ranade2007variation, benjamin2017compressed}.

Internet of Things (IoT)~\cite{lee2015internet,wortmann2015internet} is another typical application area of the SVD compression. As a crucial technique breakthrough intensively applied in areas such as transportation \cite{humayun2020}, smart building \cite{shi2020deep} and energy management \cite{al2017smart}, huge amounts of data poses enormous pressure for efficient computation and fast transmission, to solve which SVD generates a serious of efficient compression methods \cite{ZHAO2021102454, li2021random, tang2019realtime, de2015data, hashemipour2020big}. Those SVD-based compression approaches even became state-of-the-art technologies in the smart grid \cite{wen2018compression}. 

There is no doubt that SVD performs well in compression, however, with the development of detection equipments and mobile devices, the demand of data storage, data processing, data resolution as well as real-time transmission increases, which poses a more serious challenge for compression. Even though SVD as a compression tool has great significance, it inherently contains theoretical pitfalls. The basic idea of SVD is to factorize a matrix into three parts, where $\bm{U}$ and $\bm{V}$ are orthonormal matrices. Take $\bm{U}$ as an example, if it has $r$ columns, then the length of each column is one (which includes $r$ constraints), and the inner product between any two columns is zero (which includes $r(r-1)/2$ constraints). Thus \bm{$U^TU = I_r$}, where \bm{$I_r$} is the identity matrix ranked $r$. This well-known orthonormality property inherently includes $r(r+1)/2$ constraints. Therefore, for the SVD compression where both \bm{$U$} and \bm{$V$} are orthonormal matrices ranked $r$, $r(r+1)$ redundant degrees of freedom are introduced in the factorization. For the sake of clearer representation, in this article we would denote orthonormal matrices acquired by SVD as ``SVD matrices'', namely \bm{$U$} and \bm{$V$} mentioned above. Thereby, with constraints of SVD matrices removal, a deeper compression without any information loss becomes feasible.

In order to overcome the inherent deficiency, in this paper, we establish a theory to enhance SVD compression. Based on the lossless premise, this theory reduces the number of storage units of SVD matrices with constraint release and redundancy clearance and promotes the advantage of SVD compression without losing further information. Two optimized algorithms that avoid multiple matrix multiplications are presented to implement this enhancement quickly. The enhanced SVD compression is accordingly named E-SVD and results from empirical experiments again confirm its superiority over SVD. In both theory and experiment, this paper demonstrates that E-SVD losslessly enhances the SVD compression and saves as high as 25\% storage space when SVD fails to reduce any storage, while preserving the same amount of information.

\section{Theoretical Methodology} 
\label{Methodology}

The purpose of E-SVD is to extract ``freely valued elements'', elements without constraints on their values, in SVD matrices. The main theoretical tools used to implement this enhancement are Givens transformation \cite{givens1958computation} and its reverse transformation of orthonormal column matrices.

\subsection{Givens Transformation}

To facilitate the understanding, we firstly explain the basic concepts of Givens transformation.

\begin{definition}[Givens Transformation Matrix]
For a vector $\bm{W} = (w_1, w_2, ..., w_m)^T$, let $s_{ki} = \sqrt{w_k^2 + w_i^2} \ (1 \leq k<i \leq m)$, $c = cos\theta_{ki} = w_k/s_{ki}$, $d = sin\theta_{ki} = w_i/s_{ki}$, we can denote a $m \times m$ square matrix

\begin{equation}
    \textbf{$\bm{G_{ki}} =\left[
    \begin{matrix}
    \bm{I_{k-1}} && 0 && ... && 0 && 0 \\
    0 && c && ... && d && 0\\
    ... && ... && \bm{I_{i-k-1}} && ... && ... \\
    0 && -d && ... && c && 0\\
    0 && 0 && ... && 0 && \bm{I_{m-i}} 
    \end{matrix}
    \right]$}
\end{equation}as the \textbf{Givens transformation matrix} defined by the $k^{th}$ and $i^{th}$ elements of \bm{$W$}.
\end{definition}

Here \bm{$G_{ki}$} is modified from an identity matrix ranked $m$. Specifically, after replacing $(k,k)$, $(k,i)$, $(i,k)$ and $(i,i)$ elements of \bm{$I_{m}$} with $cos\theta_{ki}$, $sin\theta_{ki}$, $-sin\theta_{ki}$ and $cos\theta_{ki}$ separately, we can get \bm{$G_{ki}$}, which is consequently a orthonormal matrix of size $m$. In fact, the Givens transformation matrix in linear algebra is always used to introduce zeros in vectors or matrices with a counter-clockwise rotation, hence left multiplying this matrix accomplishes so-called Givens rotation. $\theta_{ki}$ is accordingly refer to the rotation angle of the Givens transformation, which is the only element to determine $\bm{G_{ki}}$.

\begin{lemma}
Let \bm{$G_{ki}$} $(1 \leq k<i \leq m)$ be the Givens transformation matrix defined by the $k^{th}$ and $i^{th}$ elements of vector $\bm{W} = (w_1, w_2, ..., w_m)^T$. If we denote $\bm{G_{ki}W} = (z_1, z_2, ..., z_m)^T$, then $z_k = s_{ki} = \sqrt{w_k^2 + w_i^2}$, $z_i = 0$, and $z_j = w_j \ (j \neq k,i)$.
\label{lemma1}
\end{lemma}

In particular, if we intentionally let $k = 1$, we can transform the last $(m-1)$ elements of $\bm{W}$ into zeros using a series of Givens transformation. To describe this process, denote
$$
    \bm{W^{(1)}=W}\ \text{and} \ \bm{W^{(i)}=G_{1i}W^{(i-1)}} \ (i = 2,3,...,m),
$$
where $\bm{G_{1i}}$ is the Givens transformation matrix defined by the first and the $i^{th}$ elements of \bm{$W^{(i-1)}$}. Thus repeatedly applying Lemma \ref{lemma1} we can get
\begin{equation}
    \bm{G_{1m}G_{1(m-1)}G_{1(m-2)}}...\bm{G_{13}G_{12}W} = (s, 0, 0, ..., 0)^T,
\label{zeros}
\end{equation}
where $s = \sqrt{w_{1}^2 + w_{2}^2 + ... + w_m^2}$. For convenience, we call $\bm{G_{1i}} \ (i = 2,3,...,m)$ ``the series of Givens transformation matrices defined by the first element'' of \bm{$W$}. Similarly, for every $k$, we can call $\bm{G_{ki}} \ (i = k+1,k+2, ...,m)$ ``the series of Givens transformation matrices defined by the $k^{th}$ element'' of \bm{$W$}.\par

It is quite clear that, if $\bm{W}$ is a normalized vector, namely $\bm{W^TW} = s =1$, then Equation(\ref{zeros}) becomes
\begin{equation}
    \bm{G_{1m}G_{1(m-1)}G_{1(m-2)}}...\bm{G_{13}G_{12}W} = (1, 0, 0, ..., 0)^T.
\label{0}
\end{equation}\par

Lemma \ref{lemma1} points out the result of Givens transformation for vectors, while it is also applicable to matrices. Therefore, we would propose a property beneficial for the matrix Givens transformation, which is depicted in Lemma \ref{lemma2}. 

\begin{lemma}
For any $\bm{A}$ as a $m \times r$ matrix, when left multiply it with Givens transformation matrix \bm{$G_{ki}$} $(1 \leq k < i \leq m)$, only elements in the $k^{th}$ and $i^{th}$ rows of \bm{$A$} would alter, while other elements keep unchanged.
\label{lemma2}
\end{lemma}

The proof of Lemma\ref{lemma1} and Lemma\ref{lemma2} is easily found in classical materials about matrix theory. Specifically, when denote $\bm{A} \in \mathbb{R}^{m \times r}_r$ is a matrix has full column rank $r$, the process of transforming $\bm{A}$ into $\left[
    \begin{matrix}
    \bm{R_{r\times r}}\\
    \bm{O_{(m-r) \times r}}\\
    \end{matrix}
    \right]$
using Givens transformation is also shown in most textbooks about matrix theory , where $ \bm{R_{r\times r}}$ is an upper triangle matrix ranked $r$, and $\bm{O_{(m-r) \times r}}$ is a zero matrix with $(m-r)$ rows and $r$ columns, namely all elements in $\bm{O_{(m-r) \times r}}$ are zeros.

\subsection{Givens Transformation of orthonormal Column Matrix}
Different from the classical research question in matrix theory, our object of study is not a full column rank matrix, but the ``orthonormal column matrix'': if $\bm{A} \in \mathbb{R}^{m \times r}_r \ (r<m)$, and $\bm{A^TA = I_r}$, then $\bm{A}$ is called the orthonormal column matrix. To make the Givens transformation of orthonormal column matrix clear, we propose a proposition and a theorem here, to illustrate that, a series of special Givens rotations are able to transform an orthonormal column matrix $\bm{A}$ into $\left[
    \begin{matrix}
    \bm{I_{r}}\\
    \bm{O_{(m-r) \times r}}\\
    \end{matrix}
    \right]$.

\begin{proposition}
If $\bm{A} \in \mathbb{R}^{m \times r}_{r} \ (r<m)$ is an orthonormal column matrix, let $\bm{A^{(1)} = A}$, there exists $(m-1)$ Givens transformation matrices $G_{1i}(i=2,3,...,m)$ that satisfy
\begin{equation}
    \begin{aligned}
     \bm{A^{(2)}} & = \bm{(G_{1m}G_{1(m-1)}...G_{13}G_{12})A^{(1)}} \\
     & = 
     \textbf{$\left[
    {\begin{matrix}
    1 && 0 && ... && 0 \\
    0 && a_{22}^{(2)} && ... && a_{22}^{(r)}\\
    ... && ...  && ... && ... \\
    0 && a_{m2}^{(2)} && ... && a_{mr}^{(2)} 
    \end{matrix}}
    \right]$} \\
    & =
    \textbf{$\left[
    {\begin{matrix}
    1 && 0 \\
    0 && \bm{H^{(2)}} 
    \end{matrix}}
    \right]
    $},
    \end{aligned}
\label{givens1}
\end{equation}
where $\bm{A^{(2)}}$ and $\bm{H^{(2)}}$ are also orthonormal column matrices.
\label{proposition1}
\end{proposition}

In Equation (\ref{givens1}), $\bm{G_{1i}} \ (i = 2,3,...,m)$ are the series of Givens transformation matrices defined by the first element of the first column of $\bm{A^{(1)}}$. After implementing Lemma \ref{lemma1} and Equation (\ref{0}) repeatedly, Proposition \ref{proposition1} can be obtained. Proof is shown in the Supplement Information Text 1, refer to 
which a general conclusion can be elucidated as below.

\begin{theorem}
For an orthonormal column matrix $\bm{A} \in \mathbb{R}^{m \times r}_{r}$, there exists $[(m\times r) - \frac{1}{2}r(r+1)]$ Givens transformation matrices $G_{ki}(k = 1,2,...,r; \ i= k+1,k+2,...,m)$ that satisfy
\begin{equation}
    \begin{aligned}
     &(\bm{G_{rm} G_{r(m-1)}} ... \bm{G_{r(r+1)}})...(\bm{G_{2m} G_{2(m-1)}} ... \bm{G_{23}})\\
     &(\bm{G_{1m} G_{1(m-1)}} ... \bm{G_{12}})\bm{A} = 
    \textbf{
    $\left[
    \begin{matrix}
    \bm{I_{r}}\\
    \bm{O_{(m-r) \times r}}\\
    \end{matrix}
    \right]$},
    \end{aligned}
\label{givens}
\end{equation}
where $\bm{I_r}$ is an identical matix ranked $r$ and \bm{$O_{(m-r) \times r}$} is a zero matrix with $(m-r)$ rows and $r$ columns. 
\label{theorem1}
\end{theorem}

We denote the process transforming \bm{$A$} into $\left[
    \begin{matrix}
    \bm{I_{r}}\\
    \bm{O_{(m-r) \times r}}\\
    \end{matrix}
    \right]$
as ``Givens transformation of the orthonormal column matrix''. The proof of the theorem above is shown in the Supplement Information Text 2.

\subsection{The Algorithm of Givens Transformation of orthonormal Column Matrix}

When deducing Theorem \ref{theorem1}, the computation process of Givens transforming $\bm{A} \in \mathbb{R}^{m \times r}_r$ is inherently established. However, it is worth noticing that, the dimension of every $\bm{G_{ki}}$ is $m \times m$, thus for Equation(\ref{givens}) with computation complexity O($m^4r$), it costs lots of computation resources especially when $m$ is large. Therefore, we propose a more concise algorithm to relieve the computation stress.\par

In reality, according to Lemma \ref{lemma2}, we only need to update elements in the $k^{th}$ and $i^{th}$ rows of $\bm{A^{(k)}}$ when left multiply it with $\bm{G_{ki}}$. Moreover, according to Theorem \ref{theorem1}, after the $k^{th}$ iteration, in the $k^{th}$ row of $\bm{A^{(k+1)}}$, $a^{(k+1)}_{kk} = 1$ and $a^{(k+1)}_{kj} = 0 \ (j \neq k)$, thus we can directly assign the values of elements in the $k^{th}$ row of $\bm{A^{(k+1)}}$. Due to both reasons above, when continuously left multiply $\bm{A^{(k)}}$ with $\bm{G_{ki}} (i = k+1, k+2,...,m)$, we only need to update part of the elements in the $i^{th}$ row to avoid redundant matrix multiplications.\par

Without loss of generality, in the $k^{th}(k = 1,2,...,r)$ step, whose purpose is to let all elements under the diagonal element $a_{kk}^{(k+1)}$ of \bm{$A^{(k+1)}$} to be zeros, we need to compute a series of Givens transformation matrix \bm{$G_{ki}$} $(i = k+1, k+2, ...,m)$ according to the $k^{th}$ row of $\bm{A^{(k)}}$, and update the matrix \bm{$A^{(k+1)}$} $= (a_{ij}^{(k+1)})$. The specific calculation steps are as follows:
\begin{enumerate}
    \item Denote $s_{kk} = a_{kk}^{(k)}$;
    \item For all $i = k+1, k+2,...,m$, compute $$(s_{ki})^2=\sum_{h=k}^i(a_{hk}^{(k)})^2, c_{ki} = \frac{s_{k(i-1)}}{s_{ki}}, d_{ki} = \frac{a_{ik}^{(k)}}{s_{ki}}$$ to get Givens transformation matrices \bm{$G_{ki}$}$(i = k+1, k+2, ..., m)$;
    \item To compute \bm{$A^{(k+1)}$}, for all $1 \leq h \leq k$, let 
    \begin{align}
        & a_{hh}^{(k+1)}=1, \ a_{lh}^{(k+1)} = 0 (l = h+1,...,m), \\ 
        & a_{hj}^{(k+1)} = 0(j = h+1,...,r);
    \end{align}
    \item Compute $$a_{(k+1)j}^{(k+1)} = -d_{k(k+1)}a_{kj}^{(k)} + c_{k(k+1)}a_{(k+1)j}^{(k)} \ (j = k+1,...,r);$$
    \item For all $i = k+2,...,m$, compute $$a_{ij}^{(k+1)} = -\frac{d_{ki}}{s_{k(i-1)}}\sum_{p=k}^{i-1}a_{p1}^{(k)}a_{pj}^{(k)}+c_{ki}a_{ij}^{(k)} \ (j = k+1,...,r).$$
\end{enumerate}
\par

The computation complexity of this algorithm is O($mr^2$) (generally $r \ll m$), and its pseudocode is shown in the Algorithm S1 of Supplement Information Text 3.\par
It needs to emphasize that, in the computation process above, for every Givens transformation matrices $\bm{G_{ki}}$, the only freely valued element is $\theta_{ki}$, namely $c_{ki} = cos\theta_{ki}$ and $d_{ki} = sin \theta_{ki}$. Thus for orthonormal column matrix $\bm{A} \in \mathbb{R}^{m \times r}_r$, we only need to store the corresponding $[m\times r - \frac{1}{2}r \times (r+1)]$ rotation angles $\theta_{ki}$ to obtain all Givens transformation matrices $\bm{G}_{ki}(k = 1,2,...,r; i = k+1,...,m)$. In reverse, after storing all $\bm{G}_{ki}$, we can use the reverse transformation, proposed in the next section, to get the original orthonormal column matrix.

\subsection{Reverse Transformation of orthonormal Column Matrix}
As shown in the previous section, if we have stored $[m\times r - \frac{1}{2}r \times (r+1)]$ Givens transformation matrices \bm{$G_{ki}$} $(k=1,2,\cdots,r;i=k+1,k+2,\cdots,m)$, the orthonormal column matrix \bm{$A$} can be easily recovered through
\begin{equation}
    \begin{aligned}
        \bm{A} = (\bm{G_{12}^{\textbf{T}}} \bm{G_{13}}^{\textbf{T}}...\bm{G_{1m}^{\textbf{T}}})...(\bm{G_{r(r+1)}^{\textbf{T}} 
        G_{r(r+2)}^{\textbf{T}}}...\bm{G_{rm}^{\textbf{T}}})
    {\left[ \begin{array}{c}
    \bm{I_{r}}\\
    \bm{O_{(m-r) \times r}}\\
    \end{array} 
    \right ]}.
    \end{aligned}
    \label{reverse}
\end{equation}      
We refer the process described by Equation(\ref{reverse}) to the ``reverse transformation of orthonormal column matrix''. The same problem as we discussed in Section 2.2 is that when directly implementing Equation(\ref{reverse}), matrix product would generate a large amount of redundant calculation.\par

However, considering the fact that $\bm{G_{ki}^\textbf{T}}$ is also an orthonormal matrix, and when left multiply a matrix with $\bm{G_{ki}^\textbf{T}}$, still only elements in $k^{th}$ and $i^{th}$ rows will change, thus we can simplify the reverse transformation to reduce the calculation burden.\par

Here we will build an iteration algorithm. Firstly, we denote $\bm{B} = (b_{ij}) \triangleq {\left[ \begin{array}{c}
    \bm{I_{r}}\\
    \bm{O_{(m-r) \times r}}\\
    \end{array} 
    \right ]}$, $\bm{B^{(1)}} = (b_{ij}^{(1)}) = \bm{G_{rm}^{\textbf{T}}B}$.
Then update $\bm{B = B^{(1)}}$, and continue to calculate $\bm{B^{(1)}} = \bm{G_{r(m-1)}^{\textbf{T}}B}$. By that analogy, we can apply gradually iteration according to Equation(\ref{reverse}), whose final step is $\bm{B^{(1)}} = \bm{G_{12}^{\textbf{T}}B}$. When computing $\bm{B^{(1)} = G_{ki}^{\textbf{T}}B}$, for each $j = 1,2,..,r$, there exists

\begin{equation}  
\left\{  
        \begin{aligned}
         &b_{hj}^{(1)} = b_{hj}, h \neq k,i  &\\  
         &b_{kj}^{(1)} = c_{ki}b_{kj} - d_{ki}b_{ij} &\\  
         &b_{ij}^{(1)} = d_{ki}b_{kj} + c_{ki}b_{ij} &.   
         \end{aligned}
\right.  
\end{equation}  
The computation complexity for the simplified reverse transformation is similarly O($mr^2$). The pseudocode of is presented in the Algorithm S2 of Supplement Information Text 3.

\subsection{E-SVD}

Based on the Givens transformation and its reverse transformation of orthonormal column matrices, we can easily acquire E-SVD, a method eliminating all redundancies in SVD matrices and enhancing SVD compression losslessly.\par

To this end, we would briefly review the ``SVD compression'', or ``truncated SVD''. For any $\bm{X} \in \mathbb{R}^{m \times n}_r$, after choosing $l<r$, there exists two orthonormal column matrices $\bm{U} \in \mathbb{R}^{m \times l}_l$, $\bm{V } \in \mathbb{R}^{n \times l}_l$, and a diagonal matrix $\bm{\Sigma} = diag(\sigma_1, \sigma_2, ..., \sigma_l)$ (where $\sigma_i >0, i = 1,2,...,l$) satisfy $\bm{\hat{X}} = \bm{U \Sigma V^{\textbf{T}}},$ and $\bm{\hat{X}}$ minimizes $\Vert \bm{X - \hat{X}}\Vert$. Therefore, the SVD compression can not only reduce the number of storage units, but also lose as little information $\bm{X}$ contains as possible.\par

The E-SVD we propose here, is on the basis of SVD compression, further extract corresponding rotation angles of $\bm{U}$ and $\bm{V}$ through Givens transformation, which is able to consequently delete redundant information contained in orthonormal column matrices. Actually, for $\bm{U} \in \mathbb{R}^{m \times l}_l$ and $\bm{V} \in \mathbb{R}^{n \times l}_l$, we only need $[(m - \frac{l+1}{2})\cdot l]$ and $[(n - \frac{l+1}{2})\cdot l]$ Givens transformation matrices respectively to get them according to Equation(\ref{reverse}). Those matrices are defined by their Givens rotation angles, which are the freely valued elements with less numbers than that of elements in $\bm{U}$ and $\bm{V}$. For the sake of clearer representation, we would like to denote:\par
(1) The number of storage units of \bm{$\hat{X}$} after SVD compression:\ $N\{\bm{\hat{X}}\} = (m+n+1) \cdot l$;\par
(2) The freely valued elements of orthonormal column matrix \bm{$U$}:\ $D\{\bm{U}\} = m \cdot l - \frac{1}{2} l \cdot (l+1)$;\par
(3) The freely valued elements of orthonormal column matrix \bm{$V$}:\ $D\{\bm{V}\} = n \cdot l - \frac{1}{2} l \cdot (l+1)$;\par
(4) The nonzero elements of diagonal matrix $\bm{\Sigma}$: $D\{ \bm{\Sigma} \} = l$;\par
(5) The number of storage units of \bm{$\hat{X}$} after E-SVD:\ $D\{\bm{\hat{X}}\} = D\{\bm{U}\} + D\{\bm{V}\} +D\{ \bm{\Sigma} \} = (m+n-l) \cdot l$.\par

\subsection{Storage Ratio Analysis}

\begin{figure*}[htbp]
    \centering
    \includegraphics[width=1\textwidth]{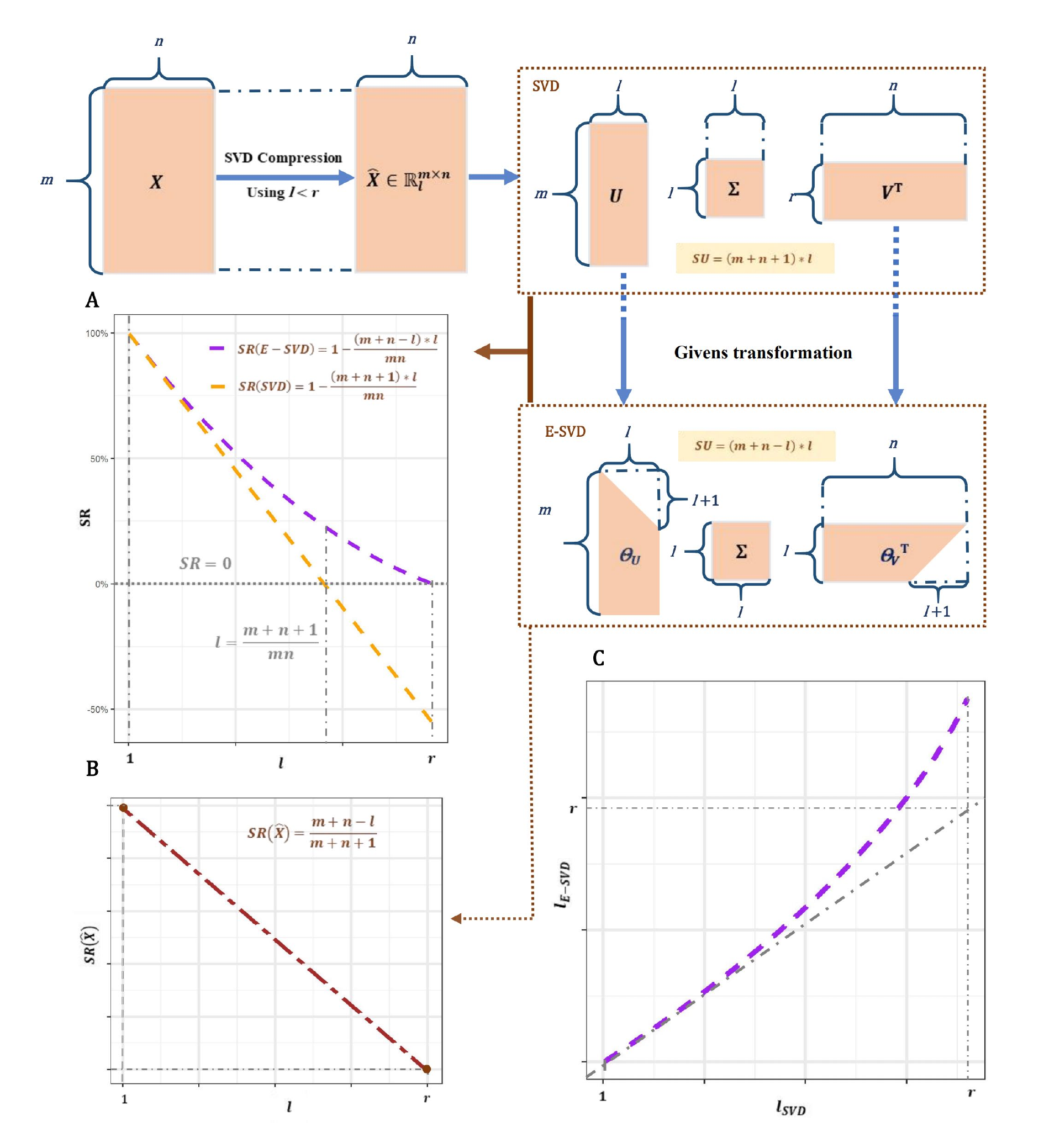}
    \caption{Storage ratio analysis and rank analysis for both E-SVD and SVD compression. A. The illustration of SR(E-SVD) and SR(SVD). B. The demonstration of SR$(\bm{\hat{X}})$ when $1 \leq l \leq r$. C. The variation of $l_\textbf{E-SVD}$ with the increase of $l_\textbf{SVD}$ based on the fixed amount of storage units, with a dash line indicating the diagonal line.}
    \label{2.3}
\end{figure*}

In order to better clarify the effect of our enhancement, we conduct this storage analysis aims at matrices before and after compression. At first, we would like to define the storage ratio (SR). In reality, many data collected with matrix forms are stored in the SVD compression format, namely it would store \bm{$U$}, \bm{$V$} as well as \bm{$\Sigma$} to get the approximation matrix \bm{$\hat{X}$}. For $\bm{\hat{X}} \in \mathbb{R}^{m \times n}_{l}$ compressed by SVD, we can use SR(SVD) to indicate the ratio of the reduced storage units to that of the original matrix, which is defined as
\begin{equation}
    \text{SR(SVD)} = 1-\frac{N\{ \bm{\hat{X}} \}}{m \cdot n} = 1-\frac{(m+n+1) \cdot l}{m \cdot n}.
\end{equation}
In a similar way, we can calculate the storage ratio when \bm{$X$} is compressed by E-SVD, which would be denoted as
\begin{equation}
 \text{SR(E-SVD)} 
  = 1-\frac{D\{ \bm{\hat{X}} \}}{m \cdot n}
  = 1-\frac{(m+n-l) \cdot l}{m \cdot n}.
\end{equation}
Also, since E-SVD is an enhanced version SVD compression, we can also illustrate the enhancement using $SR$, which is defined as
\begin{equation}
    \text{SR}(\bm{\hat{X}}) = \frac{D\{\bm{\hat{X}}\}}{N\{\bm{\hat{X}}\}} =\frac{m+n-l}{m+n+1}. 
\end{equation}
The definition domain of both SR(SVD) and SR(E-SVD) is \{$l  | l \in N_+$ and $1\leq l \leq r$\}, and higher $l$ indicates higher data fidelity, i.e, more information in the original matrix \bm{$X$} would be reserved after the compression. We can plot the variations of different $SR$ with the increase of $l$, which is shown in Fig.\ref{2.3}A. When $l=1$, SR(E-SVD)$=\frac{m+n-1}{mn} <$ SR(SVD)$=\frac{m+n+1}{mn}$. With the increase of $l$, SR of both methods decrease, whereas the advantage of E-SVD becomes more evident, since storage ratio of E-SVD is a quadratic function of $l$ as well as it is monotonically decreasing in the definition domain, thus a more remarkable compression effect compared with SVD is demonstrated. 

Another fact worth discussing is the relationship between $l$ and SR. It is obvious that for any compression technique, the basic requirement is SR $\geq 0$, whereas in Fig.\ref{2.3}A we can clearly find out SR(SVD) $<\ 0$ when $l$ approaches $r$, which means that in this situation this compression fails, whose reason is the implementation of SVD. After using the product of three matrices to approximate the original matrix, inherent constraints are included into $\bm{U}$ and $\bm{V}$, thus when $l = r$, the number of storage units of total three matrices is larger than, instead of equal to, $m \cdot n$. Nevertheless, for E-SVD, we have completely cleared up all redundancies contained in SVD matrices, therefore SR(E-SVD) $\geq 0$ always holds. Especially when \bm{$X$} has full rank, namely $r = n$ (assumes $m \geq n$), if we let $l=r$, then SR(E-SVD) $= 1-\frac{(m+n-n) \cdot n}{m \cdot n} = 0$, which means that our method needs at most $m \cdot n$ storage units. To clearer refer to the limitation of SVD compression, we would denote $l_{0}$ as the $l$ where SVD compression fails, which satisfies
\begin{equation}
    l_0 = \lfloor \frac{m \cdot n}{m
    +n+1} \rfloor.
\end{equation}

Besides, SR$(\bm{\hat{X}})$ is also a function of $l$ with a definition domain \{$l  | l \in N_+$ and $1\leq l \leq r$\}. Fig.\ref{2.3}B shows that the higher $l$ is, the lower SR$(\bm{\hat{X}})$ would be, which means that in comparison with SVD, the storage units E-SVD needs continue to decrease as the requirement for data fidelity rises. When both $m$ and $n$ are quite large and even tend to infinity, we can approximately assume $m=n$, then when $l$ reaches the limitation of SVD compression, $\lim\limits_{m \to \infty} \text{SR}(\bm{\hat{X}})$ can be calculated.

\begin{proposition}
\label{added}
For any $\bm{X} \in \mathbb{R}^{m \times n}_r$, when $m = n$ and $l = l_0 = \frac{m \cdot n}{m+n+1}$, $$\lim\limits_{m \to \infty} \text{SR}(\bm{\hat{X}}) = \lim\limits_{m \to \infty} \frac{m+n-l}{m+n+1} = \frac{3}{4} = 75 \%.$$
\end{proposition}

The proof of Proposition \ref{added} can be seen in the Supplement Information Text 4. This property is really crucial for real applications, since in extreme cases when SVD compression has already reached its limitation, E-SVD can still compress the data and use only $75\%$ storage space to preserve the same amount of information, which provides more possibilities for subsequent operations.\par

Apart from this, another critical criteria to evaluate a compression technique is the amount of information reserved. In the SVD compression, a larger $l$ means a higher similarity between $\bm{X}$ and $\bm{\hat{X}}$. 

As we discussed above, E-SVD can reserve the same amount of information with less storage space, thus with a fixed storage capacity, our method can inherently reserve more information and offer better compression quality. To prove this, we would firstly denote $l$ chosen for SVD as $l_{\text{SVD}}$, thus the number of storage units of the compressed matrix is $(m+n+1)\cdot l_{\text{SVD}}$. Then, the maximum $l$ E-SVD can get without increasing storage space is the one that satisfies
\begin{equation}
    (m+n-l) \cdot l \leq (m+n+1)\cdot l_{\text{SVD}},
\label{rprop1}
\end{equation}
which we would denote as $l_{\text{E-SVD}}$. As can be seen in Fig.\ref{2.3}C, $l_{\text{E-SVD}}$ is always higher than $l_{\text{SVD}}$ in the definition domain and the advantage over SVD keeps increasing with the rise of $l_{\text{SVD}}$.

In real applications, E-SVD can also deliver more data when the hardware capacity is limited. Here we would like to denote the maximum number of storage units for hardware as $M$, and assumes that the original matrix size is larger or equal than $M$, namely $m \cdot n \leq M$, then we can figure out the maximum $l$ we could use on the basis that every matrix is stored in SVD form. For the SVD compression, $l$ should satisfies $ M \geq \{(m+n+1)\cdot l\}$, thus the maximum $l$ corresponding to the carrying capacity, which we would denote as $l_{\text{SVD}}^{max}$, is
\begin{equation}
    l_{\text{SVD}}^{max} = \lfloor \frac{M}{m+n+1} \rfloor.
\end{equation}
Similarly, we can figure out the maximum $l$ corresponding to the carrying capacity using E-SVD, which we would denote as $l_{\text{E-SVD}}^{max}$. In our method, we need $(m+n-l)\cdot l$ units to store a compressed matrix, thus $l_{\text{E-SVD}}^{max}$ is the maximum $l$ that satisfies $M \geq \{(m+n-l) \times l\}$. According to the definition domain of $l$, it is easy to deduce that
\begin{equation}
    l_{\text{E-SVD}}^{max} = \lfloor \frac{(m+n) - \sqrt{(m+n)^2 - 4M}}{2} \rfloor.
\end{equation}

We can theoretically find out that $l_{\text{E-SVD}}^{max}>l_{\text{SVD}}^{max}$ always holds, and more important, $l_{\text{E-SVD}}^{max}/l_{\text{SVD}}^{max}$ increases fast as $M$ grows. It indicates that by fully utilizing the storage capacity, E-SVD will quickly approach to the optimal performance on information reservation. The proof is presented in Supplement Information Text 5.

\section{Empirical evidences}
\label{sec:ee}

In this section, two experiments are conducted to further provide empirical evidences of E-SVD's advantages over SVD in compression. The first one is the digital image experiment, whose purpose is to evaluate the lossless property and confirm that SR(E-SVD) is always lower than SR(SVD). Then the simulated digital matrix experiment follows, with an intention to illustrate that as the storage capacity fixed, E-SVD always obtains a higher data fidelity than SVD after the compression.\par

To evaluate the performance in compression, we will employ the mean absolute error (MAE) and the Pearson correlation coefficient ($\rho$) to compare $\bm{X}$ and $\bm{\hat{X}}$, whose dimensions are both $m \times n$, in which
$$
\text{MAE} (\bm{X}, \bm{\hat{X}}) = \frac{1}{m \times n} \sum_{i=1}^m \sum_{j=1}^n|\bm{X(i,j)} - \bm{\hat{X}(i,j)}|
$$
and
$$
\rho (\bm{X}, \bm{\hat{X}}) = \frac{\text{cov}(\bm{X},\bm{\hat{X}})}{\sigma_{\bm{X}} \sigma_{\bm{\hat{X}}}},
$$
where cov$(\bm{X},\bm{\hat{X}})$ is the covariance of vectorized $\bm{X}$ and $\bm{\hat{X}}$, $\sigma_{\bm{X}}$ and $\sigma_{\bm{\hat{X}}}$ are the standard deviations of two matrices.

\subsection{Digital Image Experiment}

We randomly chose $25$ images from a large-scale dataset for object detection in aerial images (DOTA) \cite{xia2018dota} to conduct this experiment. Note that, here, randomly selected remote sensing images are only employed to produce an intuitive and vivid picture of how effective E-SVD is in a real scenario.
Since every remote sensing image is large and has different dimension, for computation convenience, we cut each of them randomly to be $375 \times 375$. We would denote each cutted image as $\bm{X_i}\ (i = 1,2,...,25)$. For every $\bm{X_i}$, based on different $l$, we firstly approximate it using $\bm{\hat{X_i} = U_i \Sigma_i V_i^T}$ then apply E-SVD to both $\bm{U_i} \in \mathbb{R}^{375 \times l}_l$ and $\bm{V_i} \in \mathbb{R}^{375 \times l}_l$. We would denote the reconstructed matrices, i.e, the matrices after Givens transformation and its reverse transformation, as $\bm{U_i^E}$ and $\bm{V_i^E}$, and obtain $\bm{\hat{X}^E_i = U_i^E \Sigma_i (V_i^E)^T}$. We compare the difference between $\bm{U_i}$ and $\bm{U_i^E}$, $\bm{V_i}$ and $\bm{V_i^E}$ as well as $\bm{\hat{X_i}}$ and $\bm{\hat{X}^E_i}$ using MAE for 25 images separately, and all results are shown in the average form. As can be seen from Table \ref{table1}, the largest $\overline{\textbf{MAE}}$ is in the $10^{-16}$ order, which can be reasonably attribute to the machine precision. Thus the lossless property of E-SVD is confirmed numerically.

\begin{table}[htbp]
    \centering
    \begin{tabular}{| c | c | c | c |}
	\hline
	 & $\overline{\textbf{MAE}}$($\bm{U}$,$\bm{U^E}$) & $\overline{\textbf{MAE}}$($\bm{V}$,$\bm{V^E}$) & $\overline{\textbf{MAE}}$($\bm{\hat{X}}$,$\bm{\hat{X}^E}$)\\
	\hline
	$l$=50 & 5.57E-17 & 5.84E-17 & 6.39E-16 \\
	\hline
	$l$=100 & 6.61E-17 & 6.74E-17 & 6.52E-16\\
	\hline
	$l$=150 & 7.49E-17 & 7.48E-17 & 6.63E-16\\
	\hline
	$l$=200 & 8.31E-17 & 8.14E-17 & 6.71E-16\\
	\hline
	$l$=250 & 9.05E-16 & 8.76E-17 & 6.76E-16\\
	\hline
	$l$=300 & 9.73E-16 & 9.32E-17 & 6.80E-16\\
	\hline
    \end{tabular}
    \caption{The mean MAE ($\overline{\textbf{MAE}}$) of all $25$ images for $\bm{U}$ and $\bm{U^E}$, $\bm{V}$ and $\bm{V^E}$ as well as $\bm{\hat{X}}$ and $\bm{\hat{X}^E}$ in different $l$.}
    \label{table1}
\end{table}

\begin{figure*}[htbp]
    \centering
    \includegraphics[width=1\textwidth]{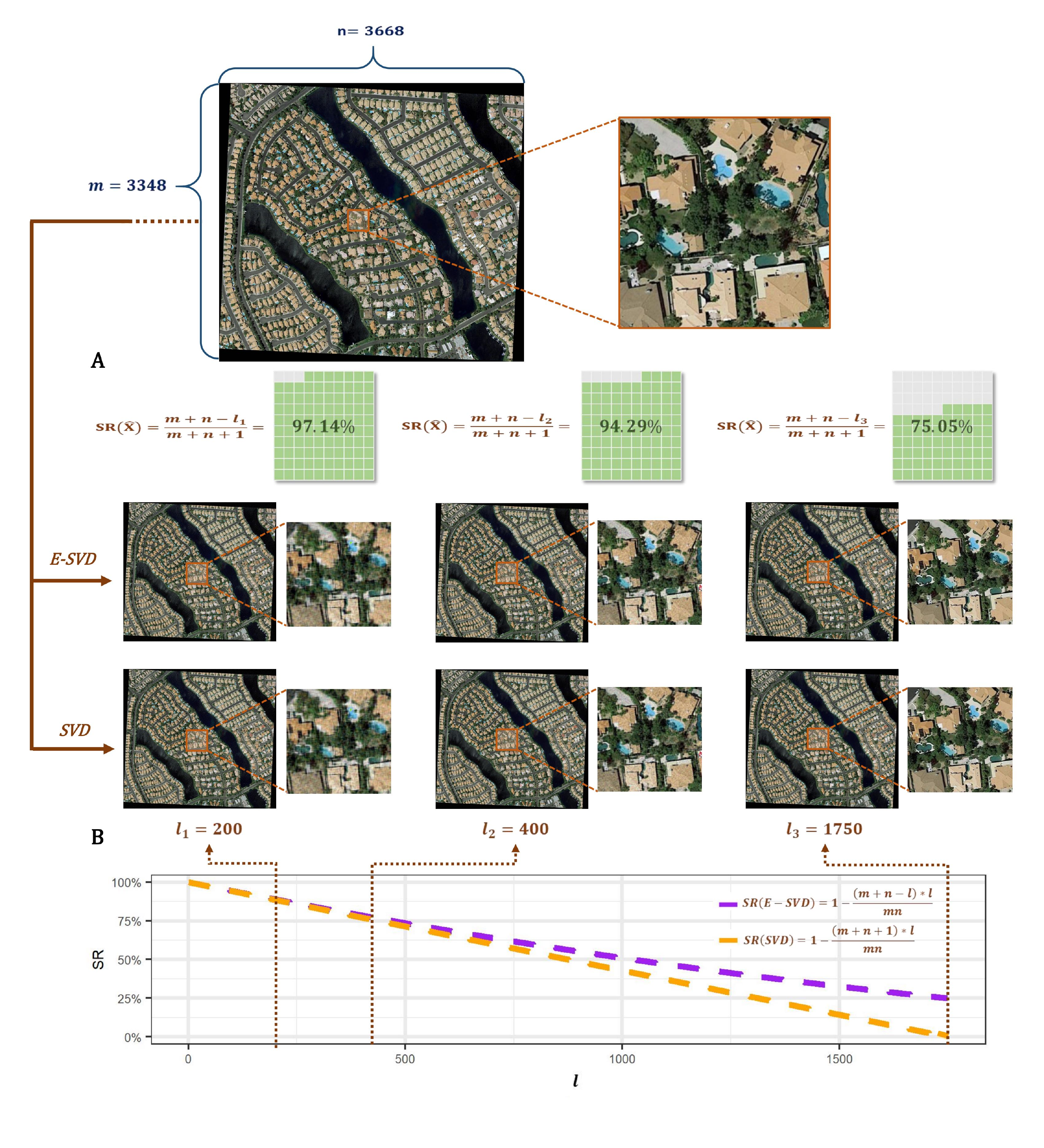}
    \caption{A. The visual clarity comparison between images compressed by SVD and E-SVD, where we applied E-SVD and SVD to red, green and blue channels of the original image respectively to get the compressed one. B. The illustration of both SR declines with increasing $l$. }
    \label{lossless}
\end{figure*}

Subsequently, we randomly chose a picture and reserve its original dimension of $3348 \times 3668$ to further illustrate the advantage of E-SVD visually as shown in Fig.\ref{lossless}. It can be easily seen that, based on different $l$, the clarity of pictures compressed by E-SVD and SVD has little distinction for naked eyes, which is in accord with numerical results, therefore our lossless property are further justified. Fig.\ref{lossless} also shows the superiority of E-SVD over SVD under the SR criteria, i.e, SR(SVD) being higher than SR(E-SVD) always holds. Moreover, With the increase of $l$, our edge over SVD increases, namely SR$(\bm{\hat{X}})$ continuous to decline. Numerically, when the image dimension is $3348 \times 3668$ and the limitation of SVD compression is $l_0 = 1750$, using E-SVD, we can use only $75.05\%$ storage units to store the same matrix $\bm{\hat{X}}$ compressed by SVD, which is close to the theoretical limitation, thus E-SVD is really significant to reduce storage space in real applications.\par

\subsection{Simulation Matrix Experiment}

Another advantage of E-SVD is its ability to store or deliver more data for the same amount of hardware capacity, thus we conduct this simulation matrix experiment to demonstrate it.

In the simulation, we let matrix $\bm{X} \in \mathbb{R}^{100\times150}_{100}$, whose elements are samples from the uniform distribution with minimum value $0$ and maximum value $100$. 
Thus the total number of storage units of \bm{$X$} before compression is $1.5 \times 10^4$. Then we would set $M$ to be a uniformly spaced sequence, whose initial value is $10^3$, end value is $1.5 \times 10^4$ and increment step is $500$. Based on different $M$, we can calculate  $l_{\text{SVD}}^{max}$ as well as $l_{\text{E-SVD}}^{max}$, which will then be used to present results.

\begin{figure*}[htbp]
    \centering
    \includegraphics[width=0.9\textwidth]{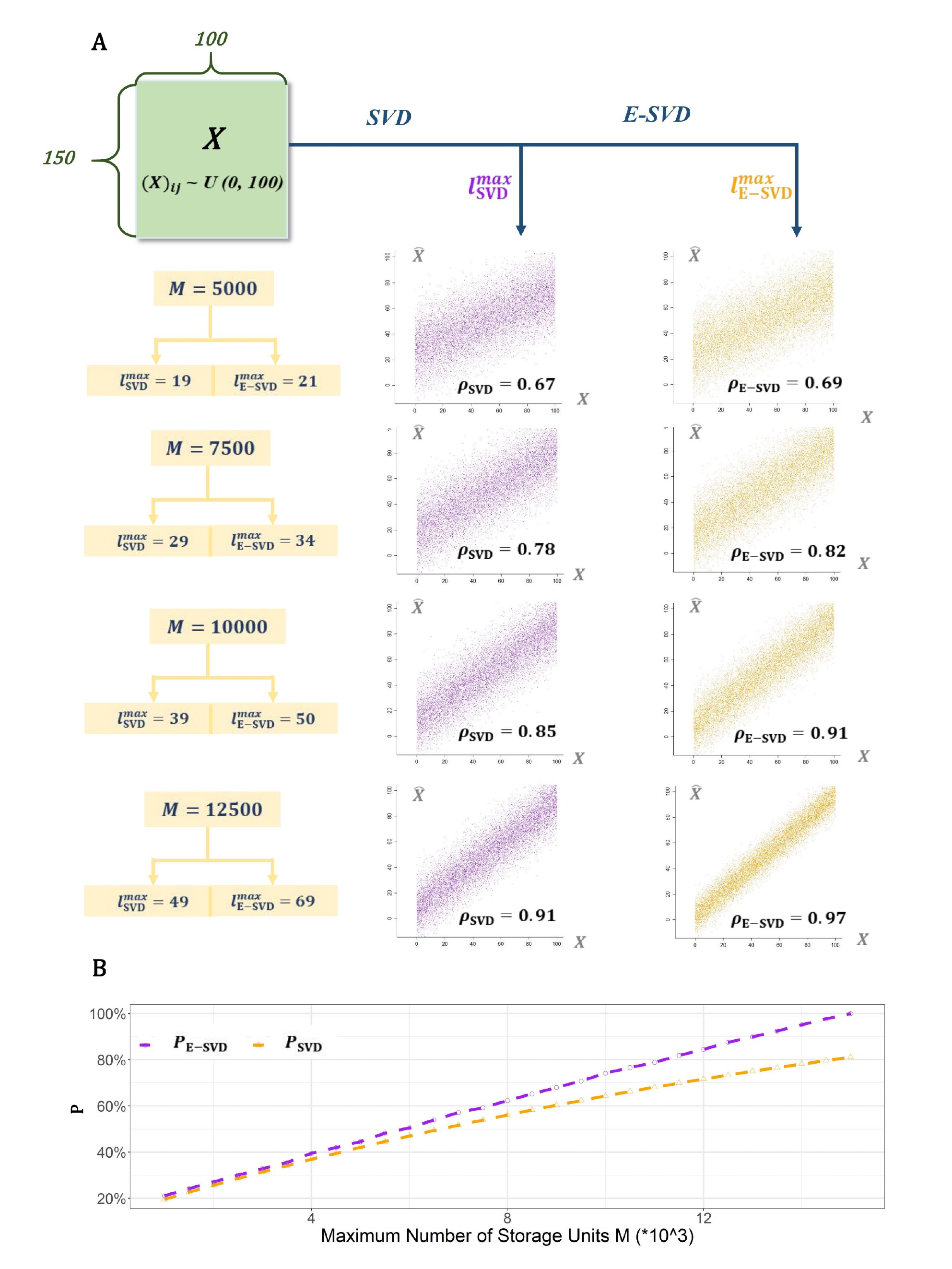}
    \caption{Correlation coefficients between a simulation matrix \bm{$X$} and its $\bm{\hat{X}}$ for $\textbf{l} = \textbf{l}_\textbf{SVD}^{max}$ and $\textbf{l} = \textbf{l}_\textbf{E-SVD}^{max}$, respectively. A. The scatter plot with $\textbf{1.5*10}^\textbf{4}$ points under different $M$, where x axis is the value of elements in $X$ and y axis is the value of elements in $\hat{\bm{X}}$. B. Information coverage percentage $\textbf{P}$ varies with increasing $M$ for both SVD and E-SVD.}
    \label{fig3}
\end{figure*}

To demonstrate the similarity between two matrices, we use the correlation coefficient $\rho$ of elements in two matrices to serve as the criteria. Here we denote $\rho_{\text{SVD}}$ as the correlation coefficient of $\bm{X}$ and $\bm{\hat{X}}$ constructed by $l_{\text{SVD}}^{max}$. Similarly, $\rho_{\text{E-SVD}}$ is the $\rho$ of $\bm{X}$ and $\bm{\hat{X}}$ constructed by $l_{\text{E-SVD}}^{max}$. We repeated our experiments 50 times to compute $\rho$ and all results are presented as averages, which are shown in Table S1 of the Supplement Information Text 6. Here we take one simulation matrix $\bm{X}$ to demonstrate our result. As shown in Fig.\ref{fig3}A, whatever the maximum storage capacity $M$ is, $\rho_{\text{SVD}} < \rho_{\text{E-SVD}}$ always holds, which indicating the persistent privilege of E-SVD over SVD. More important, the dispersion of E-SVD is clearly smaller than that of SVD, which is due to $l_{\text{SVD}}^{max} < l_{\text{E-SVD}}^{max}$, thus a greater $l$ meaning a higher similarity between \bm{$X$} and \bm{$\hat{X}$} is further illustrated.

Figure~\ref{fig3}B supports this conclusion from another perspective. To show the distinct effects, we use information coverage percentage $P$ to demonstrate the results. We define $P$ for SVD compression and E-SVD, respectively as
$$
    P_\text{SVD}(\bm{\hat{X}}) = \frac{\sum_{i=1}^{l_{\text{SVD}}^{max}}(\sigma_i)}{\sum_{i=1}^r(\sigma_i)} \
    \text{and} \
    P_\text{E-SVD}(\bm{\hat{X}}) = \frac{\sum_{i=1}^{l_{\text{E-SVD}}^{max}}(\sigma_i)}{\sum_{i=1}^r(\sigma_i)}.
$$
 We repeated our experiment 50 times to compute $P_\text{SVD}$ and $P_\text{E-SVD}$, respectively, and all results are presented as averages. As shown in Fig.\ref{fig3}B, when $M$ is small, our method has similar information coverage percentage compared with the traditional SVD. While as $M$ increases, $P_\text{E-SVD}$ keeps increasing with a high speed. It is worth noticing that when $M = 1.5 \times 10^4$, $P_\text{E-SVD}$ is $100\%$, while the traditional SVD only get $P_\text{SVD} = 81.1\%$, thus, according to this simulation, the conclusion that E-SVD clearly cleans up all redundancies is further justified, and can maximally get $18.9\%$ more information stored with the same amount of storage space as compared with SVD. 

\section{Discussion}
\label{sec:diss}

SVD is a low rank matrix factorization method widely used in areas like image compression and signal processing. For any matrix $\bm{X} \in \mathbb{R}^{m \times n}_r$, SVD can reduce the number of storage units from $(m \times n)$ to $[r \times (m+n+1)]$. As for real applications, a lower $l$ would replace $r$ to implement truncated SVD, then the number of storage units further reduce to $[l \times (m+n+1)]$. Whereas in this study, we focused on the following conclusions.

Firstly, it is well known that, when using the SVD compression, a higher data fidelity requirement always means a higher $l$. While we emphasized that the value range of $l$ has a upper bound, namely $l_0 = \lfloor \frac{m \cdot n}{m+n+1} \rfloor$. In other words, when $l_0 < l < r$, the storage units needed to store the matrix compressed by SVD would even be higher than $(m \times n)$. In this scenario, SVD fails to be a compression technique since it is no longer able to reduce the storage space. Therefore, we denote $l_0$ as the upper bound of SVD compression.

More imporatnt, due to the existence of normalization and orthogonality constraints in $\bm{U}$ and $\bm{V}$, the number of freely valued elements after SVD compression is not $[l \times (m+n+1)]$ but $[l \times (m+n+1) - l \times (l+1)]$. To this problem, on the basis of investigating Givens transformation of orthonormal column matrices, we proposed E-SVD to thoroughly eliminate redundancies in SVD decomposition, thus a deeper SVD compression is implemented losslessly. The theoretical deduction indicates that, when the number of rows and columns of a matrix is very large, and $l = l_0$ (i.e, the SVD compression fails), employing  E-SVD can still save nearly 25\% storage.

In the implementation of E-SVD, we adroitly used some special properties of Givens transformation of orthonormal column matrices to avoid a large amount of matrix multiplications. The algorithms proposed not only reduce the computation complexity, but also provide a guarantee for the practicality of E-SVD. For instance, the computation complexity of orthonormal column matrix $\bm{U} \in \mathbb{R}^{m \times l}_l$ is O($ml^2$), thus only the computation amount of the same order as one matrix multiplication is included even in the worst case, which is acceptable compared with its compression capability.

The key technique used in E-SVD is eliminating redundant degrees of freedom in orthogonormal column matrices losslessly, through which the dimension reduction process can be accomplished. In this paper, we used this technique to develop a enhanced version of SVD compression, whereas as a basic method, it has broad possible development space both theoretically and practically. In the applications of many disciplines, SVD matrices $\bm{U}$ and $\bm{V}$ is supposed to be further manipulated, such as statistical modeling or encoding the elements of matrices. However, with 
orthonormal constraints, after any operations on matrix elements, it is difficult to ensure that the resulting vectors remain standard and orthogonal. The lossless redundancy elimination technique proposed could release all constraints in orthonormal column matrices, thus an effective tool for solving problem above is provided. Apart from this, due to the wide existence of the orthonormal matrices in matrix decomposition domain, the key technique can be combined with other matrix decomposition methods, like QR decomposition\cite{gander1980algorithms}, spectral decomposition\cite{umeyama1988eigendecomposition}, Tucker decomposition\cite{kim2007nonnegative} and so on. Considering the profound role of orthonormal matrices, and the significance of matrix decomposition in a number of fields, the presented theory of losslessly removing redundancies would inspire constraint solution in more domains, thus the broad application scenarios of E-SVD is promising.

\section*{Data Availability}

The code of E-SVD has been deposited in Github at \href{https://github.com/Vicky-Zh/E-SVD}{https://github.com/Vicky-Zh/E-SVD}.

\section*{Acknowledgments}

The authors are grateful for the financial support from the National Natural Science Foundation of China (Grant Nos. 72021001 and 71871006).


\clearpage
\setcounter{algorithm}{0}
\renewcommand{\thealgorithm}{S\arabic{algorithm}}
\setcounter{table}{0}
\renewcommand{\thetable}{S\arabic{table}}
\section*{Supplement Information Text}
\setcounter{section}{0}
\section{Proof of Proposition 1.1}
\begin{proof}
 For the sake of concise statement, we would use block matrices to denote the equation.
 
 At first, if $\bm{A^{(1)}} \in \mathbb{R}^{m \times r}_r$ is a orthogonal column matrix, \bm{$G_{1i}$} $(i = 2,3,...,m)$ is the series of Givens transformation matrices defined by the first element of the first column of $\bm{A^{(1)}}$. Thus, according to Lemma 1.1,
 \begin{equation}
     \bm{A^{(2)} = G_{1m}G_{1(m-1)}...G_{12}A^{(1)}}=
      {\left[\begin{array}{cc}
     1 & \bm{\alpha_{1 \times (r-1)}} \\
     \bm{O_{(m-1) \times 1}} & \bm{H^{(2)}}\\
     \end{array}
     \right]
     },
 \label{1}
 \end{equation}
 where $\bm{O_{(m-1) \times 1}}$ is a zero matrix ranked $(m-1) \times 1$, $\bm{\alpha_{1 \times (r-1)}}$ is a matrix ranked $1 \times (r-1)$, $\bm{H^{(2)}}$ is a matrix ranked $(m-1) \times (r-1)$.\par
 
 Since $\bm{G_1 = (G_{1m} G_{1(m-1)} ... G_{13} G_{12})}$ is an orthogonal matrix, and $\bm{A^{(2)} = G_1 A^{(1)}}$, therefore
 \begin{equation}
     (\bm{A^{(2)}})^\textbf{T}(\bm{A^{(2)}}) = (\bm{A^{(1)}})^\textbf{T}\bm{G_1}^\textbf{T}\bm{G_1}(\bm{A^{(1)}}) = \bm{I_{r \times r}},
 \label{3}
 \end{equation}
 thus $\bm{A^{(2)}}$ is an orthogonal column matrix.\par
 
 Then we would like to prove that $\bm{\alpha_{1 \times (r-1)}} = \bm{O_{1 \times (r-1)}}$ and $\bm{H^{(2)}}$ is also an orthogonal column matrix. Using block matrix to denote, the left part of Equation(\ref{3}) can be written as
 \begin{equation}
 \begin{split}
     (\bm{A^{(2)}})^\textbf{T}(\bm{A^{(2)}}) 
     &= 
     {\left[\begin{array}{cc}
     1 & \bm{O_{(m-1) \times 1}^\textbf{T}} \\
     \bm{\alpha_{1 \times (r-1)}}^\textbf{T} & (\bm{H^{(2)}})^\textbf{T}\\
     \end{array}
     \right]
     }
     {\left[\begin{array}{cc}
     1 & \bm{\alpha_{1 \times (r-1)}} \\
     \bm{O_{(m-1) \times 1}} & \bm{H^{(2)}}\\
     \end{array}
     \right]
     }
     \\
     & ={\left[\begin{array}{cc}
     1 & \bm{\alpha_{1 \times (r-1)}} \\
     \bm{\alpha_{1 \times (r-1)}}^\textbf{T} & \bm{\alpha^\textbf{T}} \bm{\alpha} + (\bm{H^{(2)}})^\textbf{T}(\bm{H^{(2)}})\\
     \end{array}
     \right]
     }.
 \end{split}
 \label{4}
 \end{equation}
 At the same time, the identical matrix $\bm{I_{r \times r}}$ can be written as
 \begin{equation}
     \bm{I_{r \times r}} = 
     {\left[\begin{array}{cc}
     1 & \bm{O_{1 \times (r-1)}} \\
     \bm{O_{(r-1) \times 1}} & \bm{I_{(r-1) \times (r-1)}}\\
     \end{array}
     \right]
     }.
 \label{5}
 \end{equation}
Referring to Equation(\ref{3}), and compare elements in Equation(\ref{4}) with elements in Equation(\ref{5}), we can get
 \begin{equation}
     \bm{\alpha_{1 \times (r-1)}} = \bm{O_{1 \times (r-1)}},
 \end{equation}
 \begin{equation}
     (\bm{H^{(2)}})^\textbf{T}(\bm{H^{(2)}}) = \bm{I}_{(r-1) \times (r-1)}.
 \end{equation}
 
 Therefore, $\bm{H^{(2)}}$ is also an orthogonal column matrix. All the conclusions of Proposition 1.1 are proved.
\end{proof}

\clearpage
\section{Proof of Theorem 1.3}
\begin{proof}
 Firstly, we denote $\bm{G_1 = (G_{1m}G_{1(m-1)}...G_{13}G_{12})}$ as the product of the series of Givens transformation matrices defined by the first element of the first column of $\bm{A^{(1)} = A}$. According to Proposition 1.1, if we use the notation of block matrix, Equation(\ref{1}) becomes
 \begin{equation}
     \bm{A^{(2)}} = \bm{(G_{1m}G_{1(m-1)}...G_{13}G_{12})A} = \textbf{$\left[
    \begin{matrix}
    1 && \bm{O_{1 \times (r-1)}} \\
    \bm{O_{(m-1) \times 1}} && \bm{H^{(2)}} 
    \end{matrix}
    \right]
    $},
 \end{equation}
 where $\bm{H^{(2)}} \in \mathbb{R}^{(m-1) \times (r-1)}_{(r-1)}$, and $\bm{(H^{(2)}})^\textbf{T}(\bm{H^{(2)})} = \bm{I_{r-1}}$.\par
 
 Without loss of generality, when $1 \leq k \leq (r-1)$, we can denote the product of the series of Givens transformation matrices defined by the $k^{th}$ element of the $k^{th}$ column of $\bm{A^{(k)}}$ as $\bm{G_k = (G_{km}G_{k(m-1)}...G_{k(k+1)})}$, and $\bm{A^{(k+1)} = G_k A^{(k)}}$. Similar to the principle of Proposition 1.1, we can obtain that
 $$
     \bm{A^{(k+1)}} = \bm{(G_{km}G_{k(m-1)}...G_{k(k+1)})A^{(k)}} = \textbf{$\left[
    \begin{matrix}
    \bm{I_{k}} && \bm{O_{k \times (r-k)}} \\
    \bm{O_{(m-k) \times k}} && \bm{H^{(k+1)}} 
    \end{matrix}
    \right]
    $},
 $$
 where $\bm{H^{(k+1)}} \in \mathbb{R}^{(m-k) \times (r-k)}_{(r-k)}$, and $\bm{(H^{(k+1)}})^\textbf{T}(\bm{H^{(k+1)})} = \bm{I_{r-k}}$.\par
 
 It should be mentioned that, in the $k^{th}$ step, where $k = (r-1)$, the matrix $\bm{A^{(r)}}$ is shaped like
 $$
  \bm{A^{(r)}} = \textbf{$\left[
    \begin{matrix}
    \bm{I_{(r-1)}} && \bm{O_{(r-1) \times 1}} \\
    \bm{O_{(m-r+1) \times (r-1)}} && \bm{H^{(r)}} 
    \end{matrix}
    \right]
    $},
 $$
 where $\bm{H^{(r)}} \in \mathbb{R}^{(m-r+1) \times 1}_1$ is a column vector, and $\bm{(H^{(r)}})^\textbf{T}(\bm{H^{(r)})} = 1$.\par
 
Considering $\bm{G_r = (G_{rm}G_{r(m-1)}...G_{r(r+1)})}$ is the product of the series of Givens transformation matrices defined by the $r^{th}$ element of $\left[
    \begin{matrix}
    \bm{O_{(r-1) \times 1}}\\
    \bm{H^{(r)}}\\
    \end{matrix}
    \right]$
, we can further get
\begin{equation}
    \bm{G_r A^{(r)}} = \bm{G_rG_{r-1}...G_2G_1 A} =
    \textbf{$\left[
    \begin{matrix}
    \bm{I_{r}}\\
    \bm{O_{(m-r) \times (r)}}
    \end{matrix}
    \right]
    $}.
\end{equation}
\end{proof}

\clearpage
\section{Givens Transformation and the Reverse Transformation Algorithms}
\begin{algorithm}[htbp]
		\caption{Givens Transformation of Orthonormal Column Matrix}  
		\begin{algorithmic}[1]   
			\Require Orthogonal column matrix \bm{$A$}  
			\Ensure  Givens theta matrix \bm{$\Theta$} 
			\Function {givens\_transformation}{\bm{$A$}}   
			\State $(m,r) \gets dimension(\bm{A})$ 
			\State Create four empty matrices $\bm{C} = (c_{ij}),\ \bm{D} = (d_{ij}),\ S = (s_{ij}),\ \bm{\Theta}=(\theta_{ij})$ with $m$ rows and $r$ columns.
			\State $\bm{A^{(1)}} \gets  \bm{A}$
		    \For{each $k \in [1,r]$}\algorithmiccomment{At first, we start with computing $\theta$.}
		        \State $s_{kk} \gets a^{(k)}_{kk}$
		        \For{each $i \in [k+1,m]$}
		    	    \State $s_{ik}^2 \gets \sum_{h=k}^i(a^{(k)}_{hk})^2,\ c_{ik} \gets \frac{s_{k(i-1)}}{s_{ki}},\ d_{ik} \gets \frac{a^{(k)}_{ik}}{s_{ki}}$ 
		            \State Compute $\theta_{ik}$ subject to $sin\theta_{ik} = d$ and $cos\theta_{ik} = c$.
		       	\EndFor
		       	\State $\bm{A^{(k+1)}} \gets  \bm{A^{(k)}}, \ a^{(k+1)}_{kk} \gets 1, \ a_{(k+1)k}^{(k+1)} \gets 0$
		       	\algorithmiccomment{From here on, we begin to compute $\bm{A}^{(k+1)}$.}
		       	\For{each $j \in [k+1,r]$}
		       	    \State $a_{kj}^{(k+1)} \gets 0, \ a^{(k+1)}_{(k+1)j} \gets -d_{k(k+1)}a^{(k)}_{kj}+c_{k(k+1)}a^{(k)}_{(k+1)j}$
		       	\EndFor
		       	\For{each $i \in [k+2,m]$}
		       	\State $a^{(k+1)}_{ik} \gets 0$
		       	    \For{each $j \in [k+1,r]$}
		       	        \State $a^{(k+1)}_{ij} \gets 		       	        \frac{-d_{ki}}{s_{k(i-1)}}\sum_{h=k}^{i-1}a^{(k)}_{hk}a^{(k)}_{hj} 
		       	        +c_{ki}a^{(k)}_{ij}$
		       	    \EndFor
		        \EndFor
			\EndFor
			\State \Return{\bm{$\Theta$}}  
			\EndFunction    
		\end{algorithmic}  
	\end{algorithm}

\begin{algorithm}[htbp]
    \caption{Reverse Transformation of Orthonormal Column Matrix}
    \begin{algorithmic}[1]
        \Require Matrix dimension $m \times r$, theta matrix \bm{$\Theta$}  
		\Ensure  The original matrix \bm{$A$}
		\Function {reverse\_transformation}{$m, r, \bm{\Theta}$}
		\State Create $\bm{A^{iteration}} \in \mathbb{R}_{m \times r}^r$ where $a_{ii} = 1$ and $a_{ij} = 0$ when $i \neq j$
		\State $\bm{C} \gets cos(\bm{\Theta}), \ \bm{D} \gets sin(\bm{\Theta}), \ \bm{B} \gets \bm{A^{iteration}}, \ \bm{B^{new}} \gets \bm{B}$
		\For{each column $h$ from the $r^{th}$ column to the first column}
		    \For{each row $v$  from the $m^{th}$ row to the $(r+1)^{th}$ row}
		        \For{each $k \in [1,r]$}
		            \State $\bm{B^{new}}(h,k) =\bm{C}(v,h) * \bm{B}(h,k) - \bm{D}(v,h) * \bm{B}(v,k)$
		            \State $\bm{B^{new}}(v,k) =\bm{D}(v,h) * \bm{B}(h,k) + \bm{C}(v,h) * \bm{B}(v,k)$
		        \EndFor
		        \State $\bm{B} = \bm{B^{new}}$
		    \EndFor
		\EndFor
		$\bm{A} = \bm{B}$
		\State \Return{\bm{$A$}}
		\EndFunction
    \end{algorithmic}
\end{algorithm}

\clearpage
\section{Proof of Proposition 1.2}
\begin{proof}
 For $\bm{X} \in \mathbb{R}^{m \times n}_r$, when $m=n$ and $l = \frac{m \cdot n}{m+n+1}$, $\text{SR}(\bm{\hat{X}}) = \frac{m+n-l}{m+n+1}$, then 
 \begin{equation}
 \lim\limits_{m \to +\infty}\text{SR}(\bm{\hat{X}}) = \lim\limits_{m \to +\infty}[\frac{2m-\frac{m^2}{2m+1}}{2m+1}]=\lim\limits_{m \to +\infty}[\frac{2m}{2m+1} - \frac{m^2}{(2m+1)^2}].
 \end{equation}
 Using L'Hospital's rule, we can get
 \begin{equation}
     \lim\limits_{m \to +\infty} \text{SR}(\bm{\hat{X}}) = 1 - \frac{2m}{4(2m+1)} = 1 - \frac{1}{4} = \frac{3}{4} = 75\%. 
 \end{equation}
 Therefore, if $m = n$ and the matrix size is large enough, we can deduce that $\text{SR}(\bm{\hat{X}}) = \frac{3}{4}$ in the limitation of SVD compression.
\end{proof}

\clearpage
\section{Discussion of $l_{\text{E-SVD}}^{max}$ and $l_{\text{SVD}}^{max}$}
\label{lesvd}

 As we derived, $l_{\text{SVD}}^{max} = \frac{M}{m+n+1}$ and $l_{\text{E-SVD}}^{max} = \frac{(m+n) - \sqrt{(m+n)^2 - 4M}}{2}$, thus
 \begin{align}
     \frac{l_{\text{E-SVD}}^{max}}{l_{\text{SVD}}^{max}}
     = \frac{[(m+n)-\sqrt{(m+n)^2-4M}](m+n+1)}{2M}
     = \frac{2(m+n+1)}{(m+n) + \sqrt{(m+n)^2-4M}}.
 \label{ratio}
 \end{align}
 
 It is easy to find out that $l_{\text{E-SVD}}^{max}>l_{\text{SVD}}^{max}$ always holds, since
 \begin{equation}
     \frac{l_{\text{E-SVD}}^{max}}{l_{\text{SVD}}^{max}}
     = \frac{2(m+n+1)}{(m+n) + \sqrt{(m+n)^2-4M}}
     > \frac{2(m+n+1)}{(m+n) + \sqrt{(m+n)^2}} 
     = \frac{m+n+1}{m+n} > 1.
 \end{equation}
 Then, considering our assumption $M \leq m\cdot n$, 
 \begin{align}
     \frac{l_{\text{E-SVD}}^{max}}{l_{\text{SVD}}^{max}}
     & = \frac{2(m+n+1)}{(m+n) + \sqrt{(m+n)^2-4M}}
     \leq \frac{2(m+n+1)}{(m+n) + \sqrt{(m+n)^2-4mn}}\\
     & = \frac{2(m+n+1)}{(m+n)+|m-n|} = \frac{2(m+n+1)}{2\cdot max\{m,n\}} = 1 + \frac{min\{m,n\} + 1}{max\{m,n\}}
 \end{align}
 Therefore, when the matrix size is fixed there exists a upper bound of $l_{\text{E-SVD}}^{max}/l_{\text{SVD}}^{max}$. More interestingly, as can be seen from Equation(\ref{ratio}), with the increase of $M$, the advantage of E-SVD over SVD would becomes increasingly evident under this ratio criteria (i.e, $l_{\text{E-SVD}}^{max}/l_{\text{SVD}}^{max}$). This deduction is accordingly the theoretical guarantee of results in the simulation matrix experiment.

\clearpage
\section{Numerical results of Experiment 2}

\begin{table}[htbp]
    \centering
    \begin{tabular}{| c | c | c | c | c | c |}
	\hline
	 & \textbf{$\rho_{\text{SVD}}$} & \textbf{$\rho_{\text{E-SVD}}$} 
	 & & \textbf{$\rho_{\text{SVD}}$} & \textbf{$\rho_{\text{E-SVD}}$} \\
	\hline
	M = 1.0E3 & 0.28 & 0.33 & M = 8.5E3 & 0.81 & 0.86\\
	\hline
	M = 1.5E3 & 0.37 & 0.40 & M = 9.0E3 & 0.83 & 0.88\\
	\hline
	M = 2.0E3 & 0.43 & 0.46 & M = 9.5E3 & 0.84 & 0.90\\
	\hline
	M = 2.5E3 & 0.49 & 0.51 & M = 1.0E4 & 0.85 & 0.91\\
	\hline
	M = 3.0E3 & 0.53 & 0.55 & M = 1.05E4 & 0.87 & 0.92\\
	\hline
	M = 3.5E3 & 0.57 & 0.59 & M = 1.1E4 & 0.88 & 0.93\\
	\hline
	M = 4.0E3 & 0.60 & 0.64 & M = 1.15E4 & 0.89 & 0.95\\
	\hline
	M = 4.5E3 & 0.64 & 0.67 & M = 1.2E4 & 0.90 & 0.96\\
	\hline
	M = 5.0E3 & 0.67 & 0.69 & M = 1.25E4 & 0.91 & 0.97\\
	\hline
	M = 5.5E3 & 0.69 & 0.73 & M = 1.3E4 & 0.91 & 0.98\\
	\hline
	M = 6.0E3 & 0.72 & 0.75 & M = 1.35E4 & 0.92 & 0.98\\
	\hline
	M = 6.5E3 & 0.74 & 0.78 & M = 1.4E4 & 0.93 & 0.99\\
	\hline
	M = 7.0E3 & 0.76 & 0.80 & M = 1.45E4 & 0.94 & 1.00\\
	\hline
	M = 7.5E3 & 0.78 & 0.82 & M = 1.5E4 & 0.94 & 1.00\\
	\hline
	M = 8.0E3 & 0.79 & 0.84 & &  &\\
	\hline
    \end{tabular}
    \caption{The averaged \textbf{$\rho_{\text{SVD}}$} and \textbf{$\rho_{\text{E-SVD}}$} of all $50$ simulations for different $M$.}
\end{table}
\end{document}